\documentclass[10pt,a4paper,twoside,twocolumn,american,prd,nofootinbib,superscriptaddress]{revtex4}
\usepackage{lmodern}

\usepackage[T1]{fontenc}
\usepackage[utf8]{inputenc}
\usepackage{graphicx}
\usepackage{color}
\usepackage{babel}
\usepackage{natbib}
\usepackage{booktabs}
\usepackage{units}
\usepackage{amsmath}
\usepackage{amssymb}
\usepackage{esint}
\usepackage[normalem]{ulem}
\usepackage[unicode=true,pdfusetitle,
 bookmarks=true,bookmarksnumbered=false,bookmarksopen=false,
 breaklinks=false,pdfborder={0 0 0},backref=false,colorlinks=true]
 {hyperref}
\hypersetup{
 citecolor=blue,filecolor=blue,linkcolor=blue,urlcolor=blue}
\usepackage{subfigure} 
\usepackage{float}



\def\drm{\mathrm{d}}
\def\drdtau{\frac{\drm r}{\drm\tau}}
\def\dphidtau{\frac{\drm\varphi}{\drm\tau}}
\def\dtauds{\frac{\drm\tau}{\drm s}}

\def\dtauds{\frac{\drm \tau}{\drm s}}
\def\drds{\frac{\drm r}{\drm s}}
\def\dphids{\frac{\drm \varphi}{\drm s}}

\def\curlyh{\mathcal{H}}

\begin{document}

\title{Spherically Symmetric N-body Simulations with General Relativistic Dynamics}

\author{Julian Adamek}
\email{Julian.Adamek@unige.ch}
\affiliation{D\'epartement de Physique Th\'eorique \& Center for Astroparticle Physics,
Universit\'e de Gen\`eve, 24 Quai E.\ Ansermet, 1211 Gen\`eve 4, Switzerland}

\author{Mateja Gosenca}
\email{M.Gosenca@sussex.ac.uk}
\affiliation{Astronomy Centre, School of Mathematical and Physical Sciences, University of
Sussex, Brighton BN1 9QH, United Kingdom}

\author{Shaun Hotchkiss}
\email{S.A.Hotchkiss@sussex.ac.uk}
\affiliation{Astronomy Centre, School of Mathematical and Physical Sciences, University of
Sussex, Brighton BN1 9QH, United Kingdom}
\affiliation{Department of Physics, University of Auckland, Private Bag 92019, Auckland, New Zealand}

\begin{abstract}
Within a cosmological context, we study the behaviour of collisionless particles in the weak field approximation to General Relativity, allowing for large gradients of the fields and relativistic velocities for the particles. We consider a spherically symmetric setup such that high resolution simulations are possible with minimal computational resources. We test our formalism by comparing it to two exact solutions: the Schwarzschild solution and the Lema\^itre-Tolman-Bondi model. In order to make the comparison we consider redshifts and lensing angles of photons passing through the simulation. These are both observable quantities and hence are gauge independent. We demonstrate that our scheme is more accurate than a Newtonian scheme, correctly reproducing the leading-order post-Newtonian correction. In addition, our setup is able to handle shell-crossings, which is not possible within a fluid model. Furthermore, by introducing angular momentum, we find configurations corresponding to bound objects which may 
prove useful for numerical studies of the effects of modified gravity, dynamical dark energy models or even compact bound objects within General Relativity.
\end{abstract}
\maketitle

\section{Introduction}

Recent results from the Planck mission \cite{Ade:2015xua}, BOSS \cite{Anderson:2013zyy,Busca:2012bu,Samushia:2012iq}, WiggleZ survey \cite{Blake:2012pj}, CFHTlenS \cite{Kilbinger:2012qz} and SNLS \cite{Betoule:2014frx} have consolidated the $\Lambda$CDM concordance model of cosmology as providing a very good fit to observations. However, this model is characterized by two semi-phenomenological ingredients -- cold dark matter (CDM) and a cosmological constant ($\Lambda$) -- whose true nature still needs to be determined at the fundamental level. With the reach of linear analysis now being nearly exhausted, phenomena at nonlinear scales can help to make progress. On the observational side, large surveys such as Euclid \cite{Laureijs:2011gra}, DES \cite{Abbott:2005bi}, LSST \cite{Abell:2009aa}
and SKA \cite{Carilli:2004nx} will make significant progress analysing these non-linear scales. This puts the onus on the theoretical side to understand precisely what we expect these surveys to see. Due to the non-linearity, numerical simulations will be a necessary tool to probe this regime.

The N-body codes used for the study of cosmic large scale structure normally employ Newton's law of gravitation. One expects that this approximation works well as long as perturbations are generated by nonrelativistic matter only. This is true if dark energy is indeed a
cosmological constant and dark matter is some heavy fundamental particle (like in the WIMP scenario). However, since these facts are not established it seems that by using the Newtonian approximation we are unable to access a viable part of model space.

In fact, even within the realm of known physics this approximation will break down due to the existence of very light, but still massive, neutrinos. In principle, the initial conditions of simulations can be set late enough that neutrinos have already become non-relativistic; however, this can be so late that the cold dark matter has already begun to cluster significantly. Therefore, relativistic effects are already important in order to rigorously model the effects of neutrino masses in cosmology.

In an effort to address these shortcomings, a relativistic framework for N-body simulations has recently been developed \cite{Adamek:2013wja,Adamek:2014xba}. This framework is based on a weak-field expansion of Einstein's equations, similar to the
one proposed in \cite{Green:2010qy,Green:2011wc}. It does not require a particular form of stress-energy and relies solely on
the assumption that gravitational fields are weak, at least at large scales. Therefore, it is applicable to a much larger set of
models, including hot dark matter \cite{Davis:1992ui,Abazajian:2001nj} and many types of dynamical dark energy \cite{Copeland:2006wr}.

Before investing significant computational resources in order to do a full-scale cosmological simulation it is interesting to
study the relativistic effects in a simplified setup. Here we will consider the case of a single, isolated, spherically symmetric
structure which could, for instance, be a model for a cosmological void or a galaxy cluster. The idealization to exact spherical
symmetry drastically reduces computational requirements, allowing high-resolution simulations to be carried out at negligible cost.
Furthermore, the numerical scheme can be thoroughly verified by comparing to several known exact solutions. When comparing to exact solutions, structures can also be allowed to evolve into regimes where metric perturbations do become large and the framework breaks down, allowing us to probe the boundaries of where the framework can and cannot be trusted.

Our approach is in some sense complementary to existing methods for the numerical solution of Einstein's equations. For instance, the BSSN
formalism \cite{Shibata:1995we,Baumgarte:1998te,Rekier:2014rqa} can probe the strong field regime, but existing implementations rely on a fluid
description for matter. Our N-body method, on the other hand, allows us to study matter configurations with highly nontrivial phase space distributions.

In section~\ref{sec:model} we introduce the relativistic framework and study some simple spherically symmetric setups. We first consider
a Schwarzschild solution to confirm that the relativistic potentials are calculated accurately in vacuum. We then add nonrelativistic
matter and compare our simulations to the exact Lema\^itre-Tolman-Bondi models which describe spherically symmetric solutions with a dust
fluid. In order to avoid gauge issues, we construct several physical observables which can be compared without ambiguity. We note that
the fluid solutions break down at the formation of caustics, but our relativistic framework remains valid and can thus probe settings
beyond the fluid approximation. Without support from pressure or angular motion, overdensities tend to collapse quickly and can not
easily form stable bound objects. In section~\ref{sec:angularmomentum}, we propose a way to introduce angular motion without breaking
spherical symmetry. This is achieved by arranging the motion of the particles such that they all individually have angular momentum, but the total angular momentum of the system remains zero. We demonstrate that one can find configurations corresponding
to bound objects. Such configurations may be useful laboratories to study the effects of modified gravity, dynamical dark energy models, or even the early stages of the formation of primordial blackholes \cite{Hawking:1971ei,Carr:1974nx}, or ultra compact mini-haloes \cite{Berezinsky:2003vn,Bunn:1996da,Ricotti:2009bs}, within ordinary gravity.

\section{The Model}
\label{sec:model}

The perturbed Friedmann-Lema\^{i}tre-Robertson-Walker (FLRW) metric, in spherical coordinates and longitudinal gauge, is: 
\begin{multline}
	\drm s^2 =  -a^2(\tau)\left[1+2\Psi(\tau, r)\right] \drm\tau^2 \\
	+ a^2(\tau)\left[1-2\Phi(\tau, r)\right] \left[\drm r^2 + r^2 \drm\Omega^2\right] \, ,
\end{multline}
where $\tau$ is the conformal time, $a$ is the scale factor and we impose spherical symmetry of the perturbations by requiring that the Bardeen potentials $\Phi(\tau, r)$ and $\Psi(\tau, r)$ depend only on the radial coordinate and time. We have also assumed a spatially flat background although it would be easy to generalize our model to allow for open or closed geometries.

We examine this metric in the regime where gravitational fields are weak. In other words, we are interested in perturbations caused by structures that remain much larger than their Schwarzschild radius. Such a weak-field setting allows for a systematic expansion of the various equations of motion (including Einstein's field equations) in terms of metric perturbation variables. We will follow an approach studied in \cite{Adamek:2013wja} which takes into account the most important relativistic terms.

This approach can be summarized as follows: first, all equations are expanded in terms of the metric perturbations -- in our case $\Phi$ and $\Psi$ -- and all terms up to first order are kept without distinction. At higher orders, however, one only wants to keep the most relevant terms. Noting that linear perturbation theory is accurate on the largest scales (close to or beyond the horizon) the only higher order terms that we will keep are those which may become large at \emph{small} scales. These terms will be those with two spatial derivatives,\footnote{Note that Einstein's equations are second order differential equations, therefore no terms will have more than two derivatives.} since a derivative will effectively multiply a term by an inverse power of a length scale. To arrive at a tractable set of equations that still contains the most important relativistic corrections we will therefore add all second order terms with two spatial derivatives and no terms of any higher order. Although there are 
scenarios where terms of higher than quadratic order can dominate over the linear terms (e.g. $\Phi_{,ij}\Phi^2$ if $\Phi_{,ij}>\delta_{ij}/\Phi$) these higher order terms will always be sub-dominant to the largest \emph{quadratic} order terms that we \emph{do} include. Further  details on this approximation scheme can be found in \cite{Adamek:2013wja}.

It is important to emphasise that any perturbations of the stress-energy tensor, including momenta, are allowed to be arbitrarily large. The perturbative expansion is only carried out in terms of gravitational fields and we make no assumptions about other perturbations. For instance, our solar system perfectly fits into this scheme since the gravitational field of the sun remains well within the weak-field regime, despite the fact that its density is some thirty orders of magnitude larger than the mean density of the Universe.

Using the ``time-time'' component of Einstein's equations, $G_0^0 = 8\pi G T_0^0$, we obtain an equation for the metric perturbations: 
\begin{multline}\label{eq:Einstein1}
		\Phi_{,rr} + \frac{2}{r}\Phi_{,r} - 3 \mathcal{H} \Phi_{, \tau} - 3 \mathcal{H}^2 (\Phi -\chi) + \frac{3}{2}(\Phi_{,r})^2 \\
		= -4 \pi G a^2 (1 - 4\Phi) \delta T_0^0 \, ,
\end{multline}
where commas denote partial derivatives with respect to $r$ or $\tau$. Note that, in a spherical coordinate system as the one used here, second spatial derivatives can give rise to terms like
$\Phi_{,r} / r$. We will treat these terms like second derivatives in our expansion scheme, i.e.\ a factor $1/r$ will effectively be counted like a spatial derivative.  We also introduced the conformal Hubble parameter $\curlyh = \drm \ln a /\drm \tau $ and the difference of the potentials as $\chi = \Phi - \Psi$. On the right-hand side, $\delta T_0^0$ stands for the perturbations of the stress-energy tensor, $\delta T_0^0 = T_0^0 - \bar{T}_0^0$. We will only consider contributions from massive particles (e.g.~cold dark matter). The background model is governed by the Friedmann equation
\begin{equation}
\label{eq:Friedmann}
 \curlyh^2 = -\frac{8 \pi G}{3} a^2 \bar{T}_0^0 \, .
\end{equation}

Another equation comes from the traceless part of the ``space-space'' components of Einstein's equations, $G^i_j - \frac{1}{3} \delta^i_j G^k_k = 8\pi G \left( T^i_j - \frac{1}{3}\delta^i_j T^k_k\right)$, and reads: 
\begin{multline}\label{eq:Einstein2}
		\chi_{,rr} - \frac{1}{r}\chi_{,r}+ \chi_{,r}^2 + 2 \Phi_{,r}^2+ 2\left(\Phi_{,rr}-\frac{1}{r}\Phi_{,r}\right)(2\Phi-\chi)\\
		= 12\pi G a^2 (1 - 2\chi) \Pi_{rr} \, ,
\end{multline}
where $\Pi_{rr}$ is the radial component of the anisotropic stress, defined for a general coordinate system as:
\begin{equation}
\Pi_{ij} = \delta_{ik} T^k_j -\frac{1}{3} \delta_{ij}T^k_k \, .
\end{equation}
Latin indices denote spatial coordinates only. As a consequence of spherical symmetry the anisotropic stress is purely longitudinal in our setting.

The stress-energy tensor is derived by varying the action of an ensemble of massive point particles with respect to $\delta g_{\mu\nu}$ (see e.g. equation (2) of \cite{Adamek:2014xba}). This gives: 
\begin{multline}
T^{\mu \nu} = \sum_n m_{(n)} \frac{\delta^{(3)}(\mathbf{x} -\mathbf{x}_{(n)})}{\sqrt{-g}} \\
\times \left( -g_{\alpha \beta}\frac{\drm x^{\alpha}_{(n)}}{\drm \tau} \frac{\drm x^{\beta}_{(n)}}{\drm \tau} \right)^{-1/2} \frac{\drm x^{\mu}_{(n)}}{\drm \tau} \frac{\drm x^{\nu}_{(n)}}{\drm \tau} \, ,
\end{multline}
where we sum the contributions of $n$ particles with masses $m_{(n)}$ and spatial positions ${\bf x}_{(n)}$, and Greek indices run over all four coordinates of space-time.

For a particle moving in the radial direction we can define a momentum
\begin{equation}
\label{eq:momentum}
p = \frac{m (1-\Phi) \drdtau}{\sqrt{1+ 2\Psi -(1-2\Phi)\left(\drdtau\right)^2}} \, ,
\end{equation}
which is the proper relativistic momentum as measured in a Gaussian orthonormal coordinate frame aligned with our foliation of spacetime\footnote{Explicitly, a Gaussian orthonormal coordinate frame is given by a set of orthonormal basis vectors $e^\mu_0$, $e^\mu_1$, $e^\mu_2$, $e^\mu_3$, $g_{\mu\nu} e^\mu_0 e^\nu_0 = -1$, $g_{\mu\nu} e^\mu_0 e^\nu_i = 0$, $g_{\mu\nu} e^\mu_i e^\nu_j = \delta_{ij}$ with
$e^\mu_0$ orthogonal to the spacelike hypersurface. The metric in the coordinates defined by this basis locally looks like the Minkowski metric. The momentum $p$ defined in eq.~(\ref{eq:momentum}) is simply
the spatial component of the covariant 4-momentum in that coordinate system. If we align $e^\mu_1$ with the radial direction, i.e.\ $e^\mu_1 \propto \delta^\mu_r$, we can write $p = m u^\mu g_{\mu\nu} e^\nu_1$, where $u^\mu$ denotes the covariant 4-velocity.}. The motivation for this is that it allows us to derive an expression for the stress-energy tensor that is valid (within the bounds of our approximation scheme) even when the particles have arbitrarily high velocities. In particular,
\begin{equation}
\label{eq:T00}
T_0^0 = -\frac{1+3\Phi}{4 \pi r^2 a^3} \sum_n \delta(r - r_{(n)}) \sqrt{m^2_{(n)} + p^2_{(n)}} \, ,
\end{equation}
and 
\begin{equation}
\label{eq:Pirr}
	\Pi_{rr}  = \frac{2}{3} \frac{1+3\Phi}{4 \pi r^2 a^3}  \sum_n \delta(r - r_{(n)})
	 \frac{p^2_{(n)}}{\sqrt{m^2_{(n)}+p^2_{(n)}}}\, .
\end{equation}
As we restrict our solutions to spherical symmetry we can imagine collections of particles as representing spherically symmetric shells with only radial positions. This is because symmetry also requires that the particle distribution function is independent of angular position. We therefore simply dropped the angular coordinates from above expressions and defined the masses such that $m_{(n)} / (4 \pi r_{(n)}^2)$ is the surface mass density (in coordinate space) of the shell with label $n$. Thus, each shell accounts for all particles at given radius $r_{(n)}$ with given radial momentum $p_{(n)}$ and we need only sum over shells. Note that, for descriptive ease, from here onwards we will refer to the individual shells as the ``particles'' of our simulations.

In order to evolve the particle positions one can invert eq.~(\ref{eq:momentum}),
\begin{equation}
 \frac{\drm r}{\drm \tau} = \frac{p}{\sqrt{m^2+p^2}}(1+\Phi + \Psi) \, .
\end{equation}
The geodesic equation for massive particles,
\begin{equation}
\frac{\drm ^2 x^{\mu}}{\drm  s^2} + \Gamma^{\mu}_{\nu \rho} \frac{\drm x^{\nu}}{\drm s} \frac{\drm x^{\rho}}{\drm  s} = 0 \, ,
\end{equation}
finally determines the evolution of the momenta as
\begin{equation}\label{eq:dpdtau-noang}
\frac{\drm p}{\drm \tau} = -(\curlyh - \Phi_{,\tau}) p - \Psi_{,r}\sqrt{m^2+p^2} \, .
\end{equation}

Our aim is to study numerical solutions to the above system of equations. To this end, we adopt a particle-mesh (PM) scheme as used in many cosmological N-body simulations. The ``mesh'' part of the scheme takes care of the evolution of \textit{fields} such as $\Phi$ or $\chi$.
All fields are represented approximately by sampling their values on a discrete set of points, hereafter referred to as the ``grid''. The field equations (\ref{eq:Einstein1}), (\ref{eq:Einstein2}) are solved on the grid by replacing the differential operators by finite-difference versions thereof.

The ``particle'' part of the PM scheme, on the other hand, takes care of the evolution of the particle ensemble. The phase-space of fundamental particles is sampled by a much smaller number of N-body particles which can be viewed as discrete elements of phase-space. Hereafter, the term ``particle'' usually refers to the latter notion. Positions and momenta of particles are real-valued (i.e.\ they can exist in arbitrary positions between grid points) and the geodesic equation is solved by interpolating field-dependent quantities such as $\Psi_{,r}$ to the particle positions.

Vice versa, a so-called particle-to-mesh projection is required to construct the stress-energy tensor (whose components are treated like a field) from the particle ensemble. This is achieved by replacing $\delta(r - r_{(n)}) \rightarrow w(r - r_{(n)})$ in eqs.~(\ref{eq:T00}), (\ref{eq:Pirr}), where $w$ is a weight function which depends on the projection method.
We use the so-called ``triangular-shaped particle'' (TSP) method where $w$ is constructed using a piecewise linear (triangle-shaped, hence the name) function of the separation. Some details on the projection and interpolation methods can be found in appendix \ref{app:pm}.

In the following subsections, in order to validate the numerical scheme, we will compare simulations to two well-known exact solutions of Einstein's equations.

\subsection{The Schwarzschild Solution}

The Schwarzschild metric describes the spherically symmetric vacuum solution around a central mass concentration. Within the context of our simulations this metric is suitable for regions void of particles, and can hence be used to test the implementation of the field equations independently of the particle evolution.

In order to obtain explicit expressions for the two Bardeen potentials, it is useful to write the Schwarzschild solution in so-called ``isotropic coordinates'' \cite{Eddington:1924},
\begin{equation}
 \drm s^2 = -\frac{\left(1-\frac{r_S}{4r}\right)^2}{\left(1+\frac{r_S}{4r}\right)^2} \drm t^2 + \left(1+\frac{r_S}{4r}\right)^4 \left[\drm r^2 + r^2 \drm\Omega^2\right] \, ,
\end{equation}
where $r_S = 2 G M$ denotes the Schwarzschild radius.

As long as the mass $M$ is distributed over a central region much larger than $r_S$, the exterior Schwarzschild solution can be viewed as a perturbation around Minkowski space. Within our simulations, such a background is described by $\bar{T}_0^0 = 0$ and hence $\curlyh = 0$. We can therefore set $a = 1$ and $\tau = t$. In order to obtain a numerical solution we set up a homogeneous ball of particles
(much larger than its Schwarzschild radius) in the center of an otherwise empty simulation volume. The Bardeen potentials $\Psi$ and $\Phi$
outside of the ball are independent of time, as guaranteed by Birkhoff's theorem. Their behavior in the weak-field regime is given by the large-$r$ expansion of the above exact metric. For $r \gg r_S$ we have
\begin{subequations}
 \begin{eqnarray}
  \frac{\left(1-\frac{r_S}{4r}\right)^2}{\left(1+\frac{r_S}{4r}\right)^2} =& 1 + 2 \Psi(r) &= 1 - \frac{r_S}{r} + \frac{r_S^2}{2 r^2} + \ldots \, , \qquad\\
  \left(1+\frac{r_S}{4r}\right)^4 =& 1 - 2 \Phi(r) &= 1 + \frac{r_S}{r} + \frac{3 r_S^2}{8 r^2} + \ldots \, . \qquad
 \end{eqnarray}
\end{subequations}

\begin{figure}[tb]
 \includegraphics[width=\columnwidth]{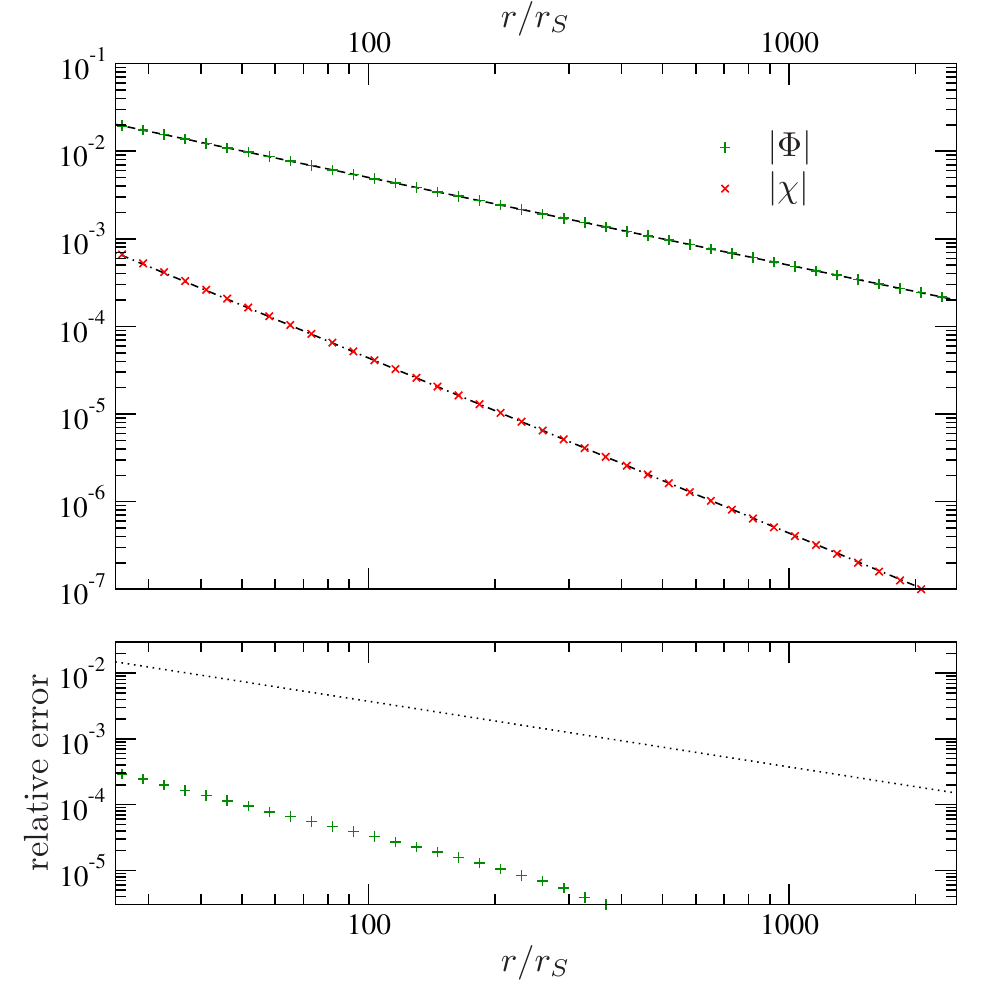}
 \caption{\label{fig:Schwarzschild} (Color online) Top: numerical results for $\Phi$ (green) and $\chi = \Phi-\Psi$ (red) as a function of $r/r_S$. They are in excellent agreement with the exact result, shown as dashed and dot-dashed lines, respectively. Bottom: relative error of the numerical values of $\Phi$ with respect to the exact result (in green). For comparison, the relative error of the Newtonian approximation, $\Phi = -G M / r$, is shown as dotted line (in black).}
\end{figure}

In Figure~\ref{fig:Schwarzschild} we show some results for $\Phi$ and $\chi$ obtained with our numerical scheme and compare them to the corresponding analytic results obtained from the exact solution. We only consider the vacuum region outside the central
mass concentration. Evidently, our numerical scheme accurately accounts for the leading-order post-Newtonian corrections and is
therefore one order (in post-Newtonian counting) better than a purely Newtonian scheme. Using the results of our simulation it would be possible, for instance, to get an accurate prediction for the advance of the peri\-helion of Mercury.

\subsection{The Lema\^itre-Tolman-Bondi Solution}

If spacetime is filled with a \emph{dust fluid} then one can construct a spherically
symmetric class of exact parametric solutions known as Lema\^itre-Tolman-Bondi (LTB) models. For our simulation this is equivalent to requiring that particles occupying identical space-time points also have identical velocities, or more precisely, having a phase space distributon function which, at each spacetime point, is a single Dirac delta-distribution in velocity space.
Being exact solutions, these models do
not require the density to be nearly homogeneous, allowing the study of strongly non-perturbative settings. However, since the
construction makes use of a comoving synchronous coordinate system in an essential way, it is impossible to extend these solutions
beyond the point where particle trajectories cross and the phase space distribution function loses its simple delta-distribution
character. In this coordinate system, the LTB line element reads

\begin{equation}
\label{eq:LTBmetric}
 \drm s^2 = -\drm t^2 + \frac{\left[R_{,r}(t,r)\right]^2}{1 + 2 E(r)} \drm r^2 + R^2(t,r) \drm\Omega^2 \, .
\end{equation}

In order to compare the LTB solutions to our numerical calculations we choose initial conditions such that
the density perturbation is linear. In the perturbative regime
we can easily work out the gauge transformation between the synchronous comoving coordinates used to parametrize the LTB
solution and the coordinates used in our framework, which are related to the longitudinal gauge. Details can be found in
appendix~\ref{app:gauge}. Setting identical initial conditions in this way we can make a comparison based on identical physical situations.

\begin{figure*}[tb]
\centering
\includegraphics[width=\textwidth]{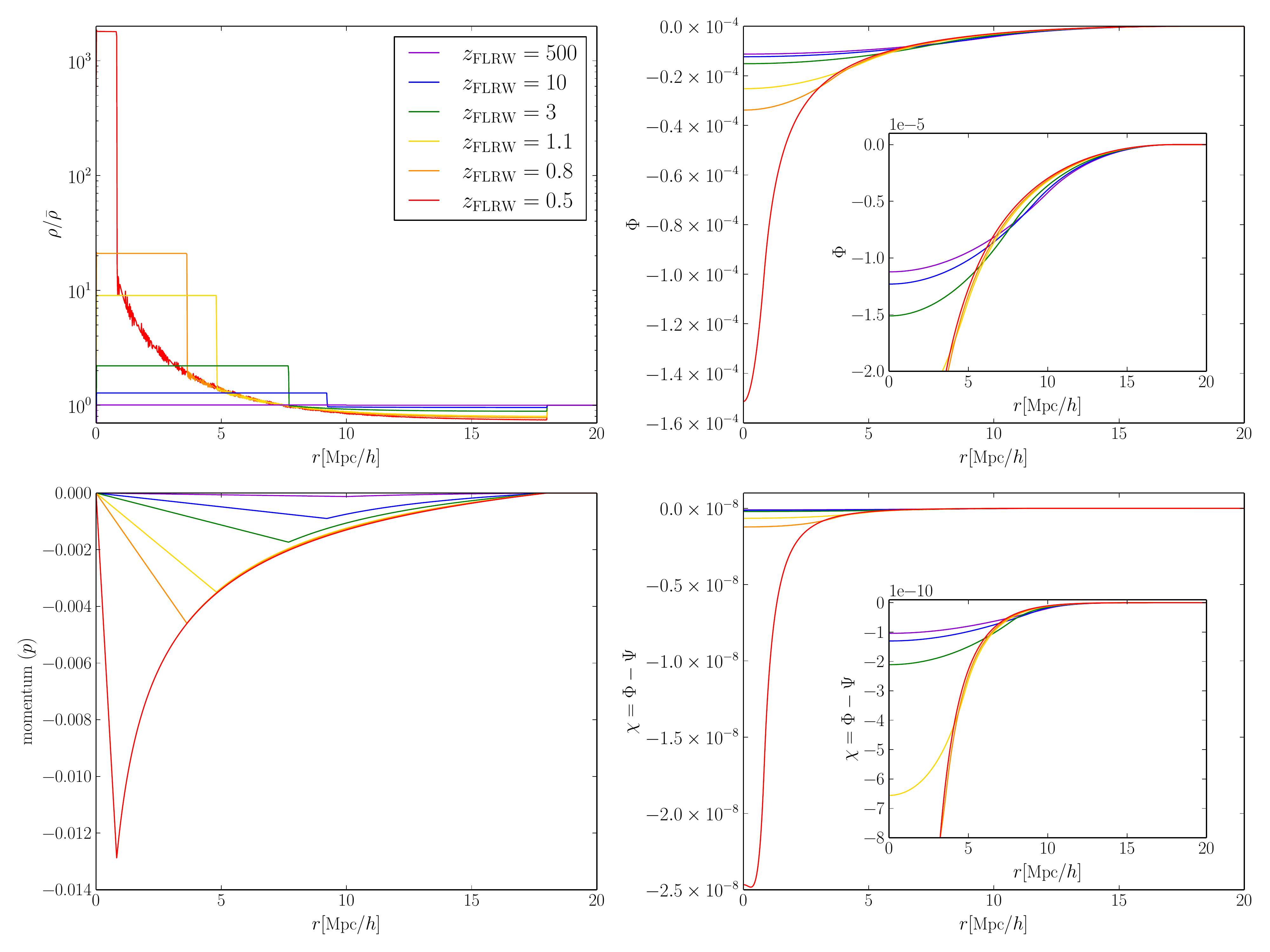}
\caption{\label{fig:overdensity} (Color online) Top left: the evolution of the density profile of a spherically symmetric compensated tophat perturbation. Different-coloured lines correspond to outputs at different times in the simulation, parametrised by the background redshift $z_{\mathrm{FLRW}}$. The last output at $z_{\mathrm{FLRW}}=0.5$ happens just before the collapse occurs. Density plots exhibit some discreteness noise, which is caused by having a finite number of particles. Bottom left: the momentum of the shells moving inwards. Top right: evolution of the underlying scalar metric perturbation $\Phi$. This profile is continuous even where the density has a step. Bottom right: Evolution of the difference of the two potentials $\chi = \Phi - \Psi$. This is a purely relativistic quantity and does not exist in a Newtonian setup.
The magnitude of $\chi$ is $\sim\Phi^2$. Parameters of the simulation were: size of the box: $20$ Mpc$/h$, initial radii of top-hat overdensity and compensated region, respectively: $r_1=6$ Mpc$/h$, $r_2=18$ Mpc$/h$, initial density contrast of the overdensity: $\delta=1/200$, initial redshift: $z_{in}=500$.}
\end{figure*}

\begin{figure*}[tb]
\centering
\includegraphics[width=\textwidth]{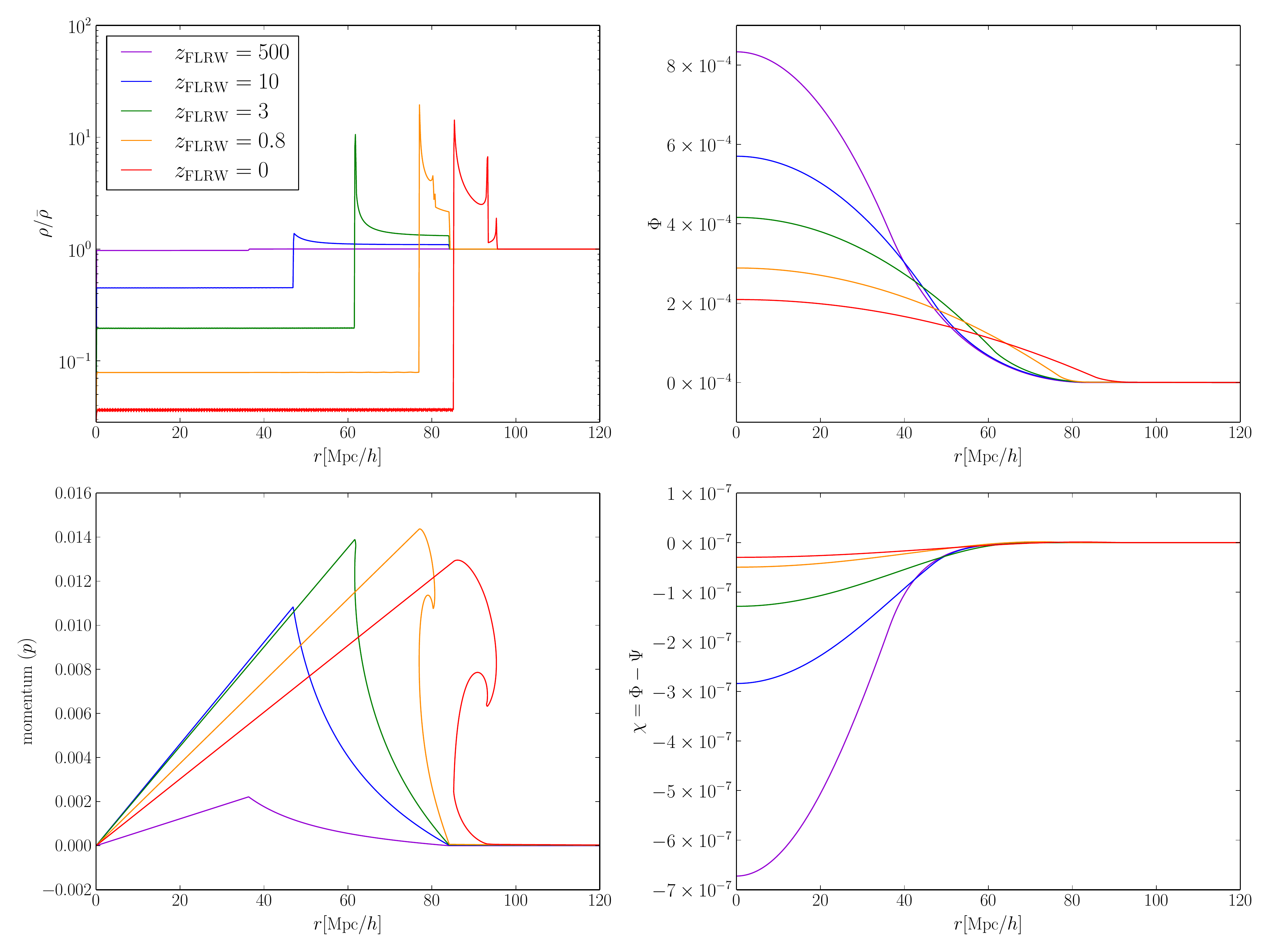}
\caption{\label{fig:underdensity} (Color online) The same set of plots as for figure \ref{fig:overdensity}, this time for an evolving underdensity. The last two outputs exhibit a shell-crossing, which can be seen in the density and momentum portraits. This non-linear feature can only be modelled with an N-body simulation. The parameters used here were: size of the box: $120$ Mpc$/h$,  $r_1=36$ Mpc$/h$, $r_2=84$ Mpc$/h$, $\delta=-1/40$, $z_{in}=500$.} 
\end{figure*}

Figures \ref{fig:overdensity} and \ref{fig:underdensity} show some simulation results. We plotted the evolution of the density contrast, momentum of particles $p$, scalar perturbation $\Phi$ and the difference of the two potentials $\chi$ as a function of comoving radius $r$. The first set of figures portrays the collapse of a spherically symmetric overdensity and the second set shows the
expansion of a spherically symmetric void. The initial densities were set as ``compensated tophat profiles'', where a central region of
constant density contrast is surrounded by a second layer with constant density contrast of opposite sign such that the entire region can be
matched onto a homogeneous FLRW exterior solution. In both cases we find $\chi$ to be proportional to $\sim \Phi^2$ and negative. 

In the expansion of a void, we can also observe a ``shell-crossing'' which happens when a set of particles moves outwards faster than particles that were initially at a larger radius. As mentioned before, this cannot be represented with an LTB solution -- it becomes singular as soon as shell crossing occurs.

When following the respective solutions into the non-linear regime we face the problem that the gauge transformation
becomes highly nontrivial. This makes it difficult to compare quantities like the density or metric components directly, because they are gauge-dependent. A good way to proceed is to compute \textit{observables}, which are by definition gauge-independent.
In the following section we will discuss some examples in detail.

\subsection{Observables}

\subsubsection{Redshift of radially in-falling source}
The first observable we will study is the redshift of a source of light that is moving with the flow of particles surrounding it. We place this source of light at an initial radius $r_1$ from the center and an observer at $r_2$, the boundary where the inhomogeneous LTB
patch is matched to FLRW. For this example, the source and observer are along the same radial line. The source constantly emits photons at a fixed energy given in the rest frame of the source. 
As the simulation progresses, we propagate these photons through the simulation volume until they reach the observer. There they are detected and we calculate their observed redshift, which is defined as: 
\begin{equation}
	1+z_\mathrm{obs} =\frac{(g_{\mu\nu} k^{\mu} u^{\nu})|_{\mathrm{src}}}{(g_{\mu\nu} k^{\mu} u^{\nu})|_{\mathrm{obs}}} \, ,
\end{equation}
where the product of the photon's 4-momentum $k^{\mu}$ and the 4-velocity of the source (observer) $u^{\mu}$ can be related
to the momentum $p$ of the source (observer) particle in the limit of weak fields as:	
\begin{equation}\label{eq:product_k_u}
	g_{\mu\nu} k^{\mu} u^{\nu} = -k^0 a \left(1+\Psi\right)\frac{1}{m}\left[\sqrt{m^2+p^2} - p\right] \, .
\end{equation}
To propagate a photon through the simulation, we use the null condition ($ds^2=0$). We can actually find a fixed relation between $dr/d\tau$ and $d\varphi/d\tau$, a consequence of the fact that a photon always travels at the speed of light:
\begin{equation}
\label{eq:nullcondition}
 1 + 2 \Psi + 2 \Phi = \left(\frac{\drm r}{\drm\tau}\right)^2 + \left(\frac{\drm\varphi}{\drm\tau}\right)^2 r^2
\end{equation}
which gives: 
\begin{equation}
\frac{\drm r}{\drm\tau} = \pm \left(1 + \Psi + \Phi\right) \qquad \Leftrightarrow \qquad \frac{\drm \varphi}{\drm \tau} = 0.
\end{equation}
On every step, the photon's energy can be evaluated by integrating the time component of the geodesic equation:
\begin{equation}\label{eq:photon_energy}
\frac{\drm k^{0}}{\drm \tau} + \left[ \Psi,_{\tau} -\Phi,_{\tau} + 2\Psi,_{r} \frac{\drm r}{\drm \tau}+ 2 \mathcal{H} \right] k^{0} =0 
\end{equation}

\begin{figure}[tb]
\includegraphics[width=\columnwidth]{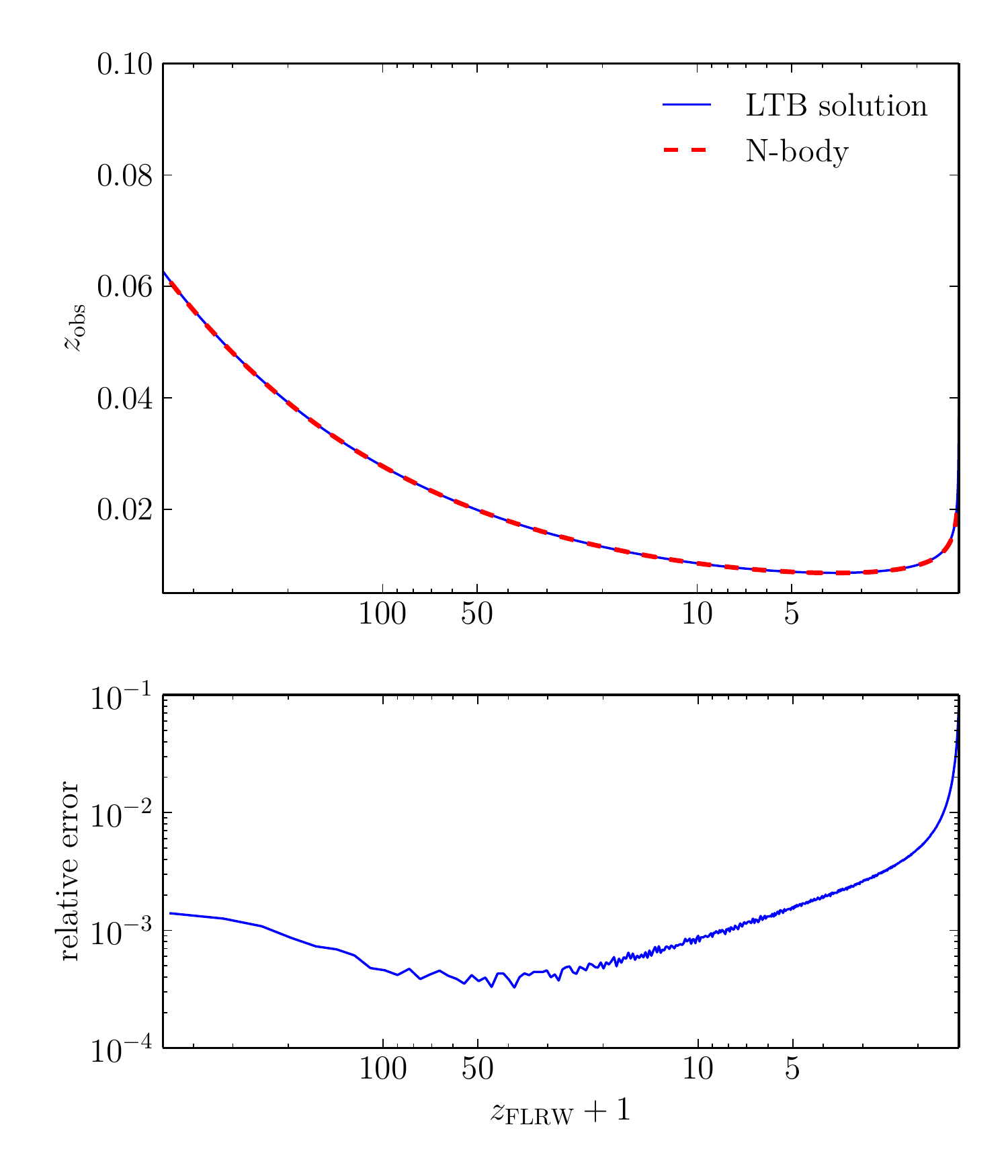}
\caption{\label{fig:rs-compare} (Color online) Top: the redshift of light, emitted by an in-falling source particle, as a function of the background redshift $z_{\mathrm{FLRW}}$ at the time when the light is detected by the observer. Initially, the observed redshift is decreasing, which is something we would expect for two particles comoving with the background in a matter dominated universe. Later, as the collapsing structure evolves, the velocity of the in-falling source becomes the dominant contribution and the redshift starts increasing again.
Bottom: relative error between our relativistic simulation and the LTB solution. The error is mainly due to the first-order matching of the initial conditions. The error increases as the collapse evolves. This is because the collapse time itself receives a first-order correction. Therefore, the divergence in observed redshift happens at slightly offset times, resulting in the error blowing up. For this plot, the same parameters as the ones in Figure \ref{fig:overdensity} were used.}
\end{figure}

The results for this observable are shown in figure \ref{fig:rs-compare}. As can be seen, the simulation agrees well with the LTB predictions. In fact, for this plot the leading source of deviation actually comes from imprecise matching of initial conditions, which could only be improved by performing that matching at a higher order of perturbation theory.

\subsubsection{Lensing of non-radial rays}

Another observable we can analyse is the deflection of a ray that propagates not only in the radial, but also in an angular direction. The trajectory of such a ray is lensed by the gravitational potentials. Spherical symmetry ensures that the trajectory of a light ray will be planar, so we only have to consider the radial direction and one angular direction. By setting $\vartheta=\pi/2$ or $k^{\vartheta}=0$, one can derive the two equations that determine that path of the photon from the geodesic equation:
 \begin{subequations}\label{eq:photon_path}
\begin{multline}\label{eq:photon_path_1}
\frac{\drm^2 r } {\drm\tau^2} 
- 2 \left(\Psi,_{r} + \Phi,_{r}\right) \left(\drdtau\right)^2 
- \left(\Psi,_{\tau} +\Phi,_{\tau}\right)\drdtau \\
- \left(\dphidtau\right)^2 r +(\Phi,_{r}+\Psi,_{r}) 
=0
\end{multline}
\begin{multline}\label{eq:photon_path_2}
 \frac{\drm^2 \varphi } {\drm\tau^2} 
 + \left(\Phi,_{\tau} -\Psi,_{\tau} - 2 \Phi,_{\tau} \right) \dphidtau\\
 + \left(\frac{2}{r} -2 \Psi,_{r} -2 \Phi,_{r}\right) \drdtau \dphidtau 
 = 0
\end{multline}
\end{subequations}
Varying the initial angle at which photons enter the perturbed region and observing the deflection angle (the angle by which its outgoing trajectory differs from the incoming one), we get a gauge independent probe of the underlying potentials. 

\begin{figure}[tb]
\includegraphics[width=\columnwidth]{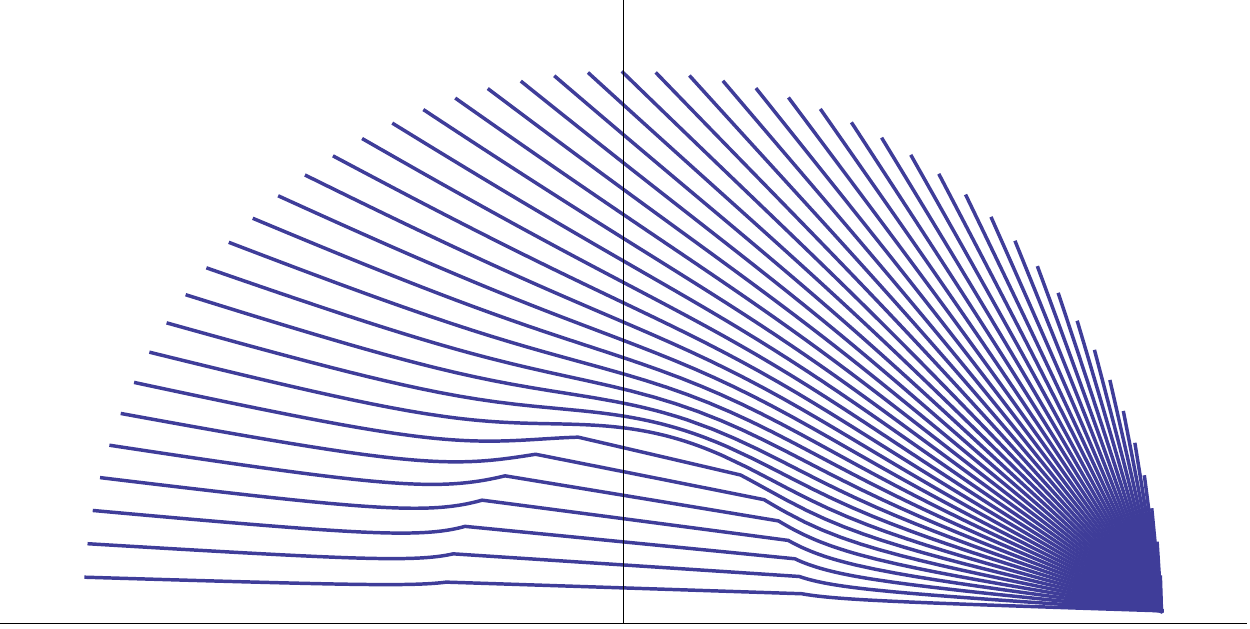}
\caption{\label{fig:lensing1} Schematic representation of the trajectories of lensed photons in an LTB geometry. The rays enter the LTB region at varying angles on the right end of the plot. Along their way through the simulation volume their trajectories are deflected due to the underlying overdensity.}
\end{figure}

\begin{figure}[tb]
\includegraphics[width=\columnwidth]{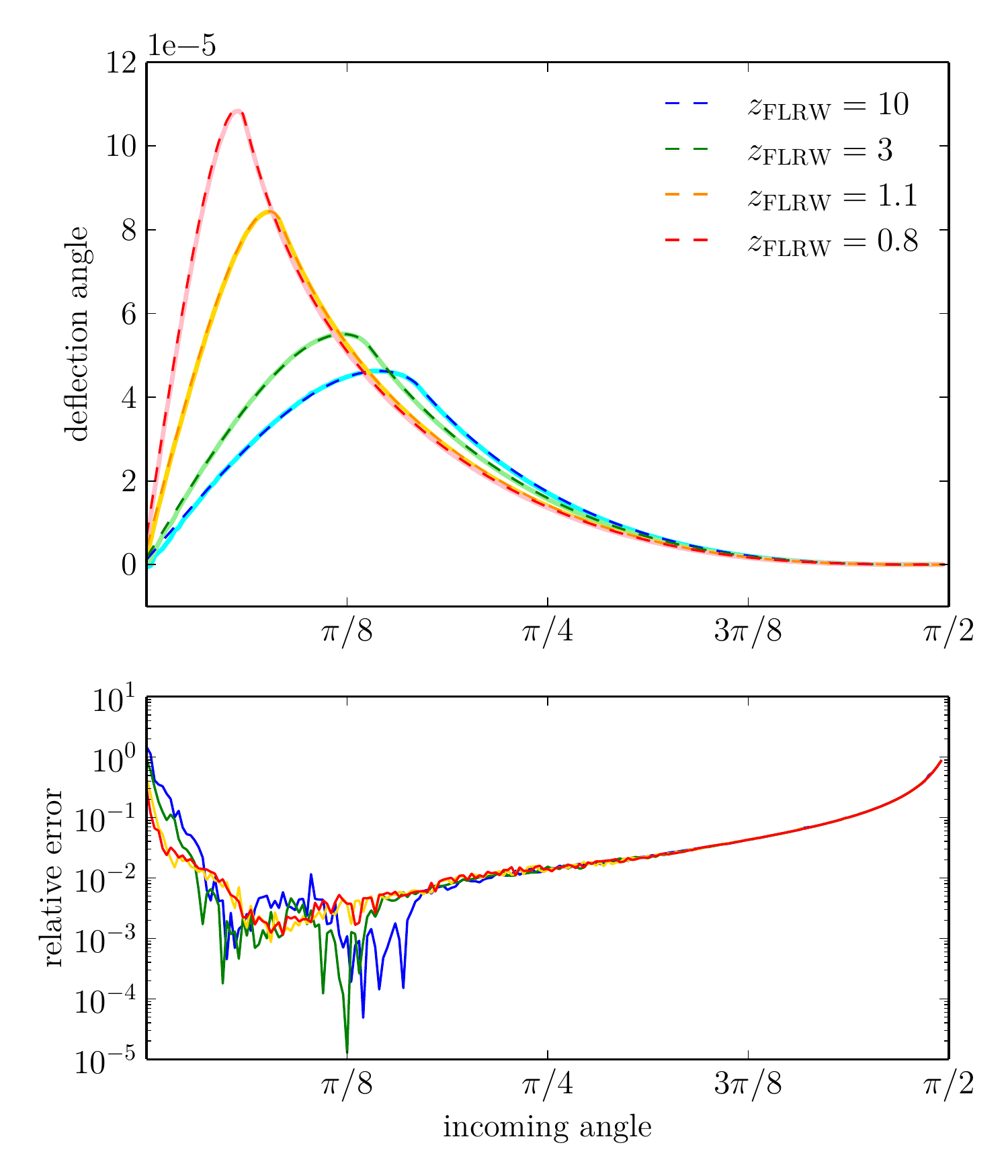}
\caption{\label{fig:lensing2} (Color online) Lensing of photons by an underlying overdensity. We plot the LTB predictions (solid lines) and the results from our simulation (dashed lines on top of them) at four selected redshifts. The parameters used here are the same as the ones in Figure \ref{fig:overdensity}.}
\end{figure}

In Figure \ref{fig:lensing1} we show the trajectories of photons that propagate through the simulation volume. We tracked $200$ photons, entering at different angles $\alpha$. The photons that experience the most lensing are those that pass near the edge of the overdensity. A photon that enters at $\alpha =0$ and passes through the centre of the overdensity is not lensed, since its path is radial. Likewise, a photon that enters at $\alpha=\pi/2$ spends too little time inside the non homogeneous region to change its path substantially. In Figure \ref{fig:lensing2} we show the deflection angle as a function of $\alpha$.

\section{Angular Momentum}
\label{sec:angularmomentum}

Using the formalism presented so far, any initial over-density would collapse to a singularity within a finite amount of time. For realistic cosmological structures this does not happen due to the process of virialisation during which the initial potential energy of a large scale density fluctuation is partially converted into radial and angular kinetic energies of individual particles. Most importantly, the angular momentum thus generated in individual particles causes these particles to miss the centre of a collapsing structure. This then avoids the production of densities large enough to cause singularities to arise. Essentially, the produced angular momentum provides an effective pressure term that resists the collapse.

We show in this section how we can model this pressure by adding angular momentum to the particles in our simulation box, without losing spherical symmetry. The downside to the method we present is that there can be no exchange of angular momentum between particles. This is because adding such an effect would necessarily require some degree of deviation from spherical symmetry. Unfortunately, this limits how far we can model the true virialisation process; nevertheless, as we show, we can still set initial conditions that produce stable, bound, spherical structures.

We can now imagine the shells to be made up of infinitesimal point-like particles, evenly distributed over the sphere. Apart from the radial momentum, which is the same for all infinitesimal point-like particles on a given sphere, we can assign each of those particles an angular momentum in a particular direction, in such a manner that once we perform the average over the momenta of all infinitesimal particles on a given sphere,
there is no preferred direction of angular momentum. One can imagine that for every infinitesimal particle with some angular momentum, there is another particle on a trajectory in the same plane, but traveling in the opposite direction. Although we can not observe any non-radial motion of spherical shells, their radial motion is nevertheless affected. This is because the equation that governs the propagation of particles in the radial direction involves a pressure-like term that depends on the angular momentum of the particle.

\begin{figure*}[tb]
\centering
\includegraphics[width=\textwidth]{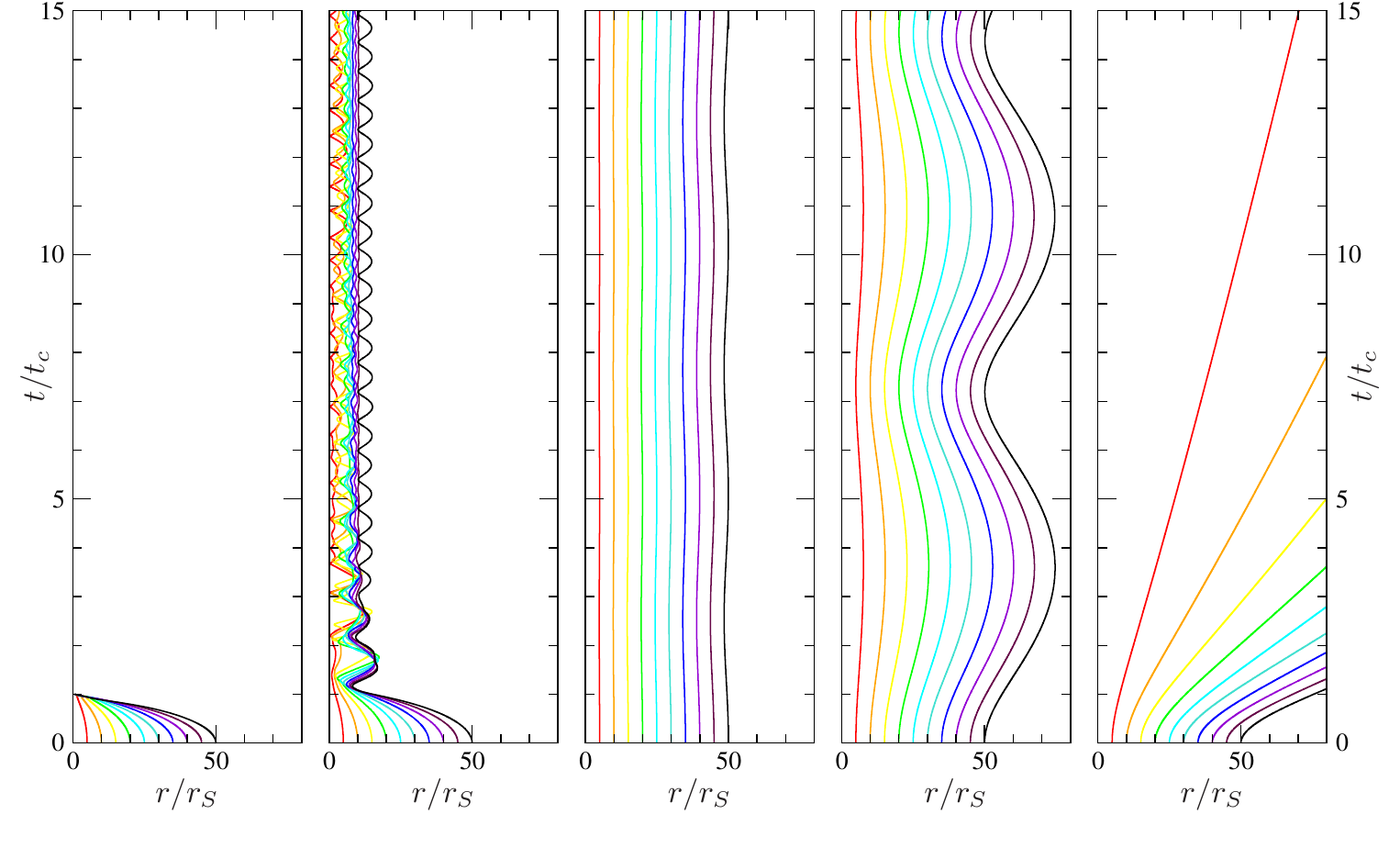}
\caption{\label{fig:orbits} (Color online) Space-time diagrams showing the radial trajectories of shells initially arranged as a uniform ball with zero radial motion. Different colors label different
shells, black labelling the outermost one. From left to right, the angular momentum of the shell particles was set to $L = 0$ where we get collapse; $L= 0.5 L_\mathrm{K}$ where a violent re-configuration to a new, much more compact state happens; $L=L_\mathrm{K}$ at which point the behaviour is almost Keplerian, with tiny perturbations caused by relativistic corrections - stability is guaranteed for a very long time scale; $L=1.1 L_\mathrm{K}$ with quasi-stable oscillations with relatively long timescale for chaotic 
behavior; and $L=1.4 L_\mathrm{K}$ which corresponds to an unbound state.
Here, $L_\mathrm{K}$ is the ($r$-dependent) value of $L$ which corresponds to a circular Keplerian orbit in Newtonian theory. The radius is plotted in units of $r_S$, the Schwarzschild radius of the ball; initially the ball has a radius of $50 r_S$. The time coordinate is plotted in units of $t_c$, the collapse time of the irrotational ball. }
\end{figure*}

We start again with the geodesic equation 
\begin{equation}
\frac{\drm ^2 x^{\mu}_{(n)}}{\drm  s^2} + \Gamma^{\mu}_{\nu \rho} \frac{\drm x^{\nu}_{(n)}}{\drm s} \frac{\drm x^{\rho}_{(n)}}{\drm  s} =0 
\end{equation}
Calculating all the Christoffel symbols $\Gamma^{\mu}_{\rho \sigma}$ we have three equations for evolution of coordinates:
\begin{subequations}
\begin{multline}\label{eq:d2ts}
\frac{\drm^2\tau}{\drm s^2} =  -\left(\curlyh + \Psi_{,\tau}\right)\left(\dtauds\right)^2 - 2 \Psi_{,r} \dtauds \drds \\
- \left(\curlyh(1-2\Phi -2\Psi) - \Phi_{,\tau}\right)\left( \left(\drds\right)^2 +r^2 \left(\dphids\right)^2 \right) 
\end{multline}
\begin{equation}\label{eq:d2rs}
\begin{split}
\frac{\drm^2r}{\drm s^2} = & - \Psi_{,r} \left(\dtauds\right)^2 - 2\left(\curlyh - \Phi_{,\tau}\right)\dtauds \drds\\
& + \Phi_{,r} \left(\drds\right)^2  + r \left(1 - r \Phi_{,r}\right)\left(\dphids\right)^2 
\end{split}
\end{equation}
\begin{equation}\label{eq:d2ps}
\frac{\drm^2\varphi}{\drm s^2} = -2  \left(\curlyh - \Phi_{,\tau}\right)\dtauds \dphids  -2   \left( \frac{1}{r}  - \Phi_{,r}\right) \drds \dphids
\end{equation}
\end{subequations}
Here we only kept one angular coordinate. Since the potentials are spherically symmetric, each infinitesimal point-like particle will move in a two-dimensional plane and so the reference frame can always be rotated so that the direction of the angular momentum is perpendicular to this plane. We see that the last equation (\ref{eq:d2ps}) can be integrated analytically with respect to $\drm s$ once, to give
\begin{equation}\label{eq:dphids}
\dphids = \frac{L}{(a r)^2} (1+ 2\Phi)
\end{equation}
where $L$ is the constant of integration. 
We can express the evolution of coordinates $r$ and $\varphi$ with respect to coordinate time $\drm \tau$ instead of eigentime $\drm s$ using a simple trick:
\begin{equation} 
 \frac{\drm^2r}{\drm\tau^2}  = \left( \frac{\drm^2r}{\drm s^2} - \frac{\drm^2\tau}{\drm s^2} \frac{\drm r}{\drm\tau}\right) \left(\frac{\drm\tau}{\drm s}\right)^{-2} 
\end{equation}
And equivalently for $\varphi$. With this we can combine equations (\ref{eq:d2rs}) and (\ref{eq:d2ps}) with (\ref{eq:d2ts}) to express: 
\begin{widetext}
\begin{subequations}
\begin{equation}
\begin{split}
\frac{\drm^2r}{\drm\tau^2} =& \left( -\curlyh + 2 \Phi_{,\tau} + \Psi_{,\tau} \right) \drdtau - \Psi_{,r} + \left(\Phi_{,r} + 2\Psi_{,r}\right)\left(\drdtau\right)^2 + \left(r - r^2 \Phi_{,r}\right)\left(\dphidtau\right)^2\\ 
&+ \left(\curlyh (1-2\Phi -2 \Psi ) -\Phi_{,\tau}\right) \left( \left(\drdtau\right)^2 + r^2 \left(\dphidtau\right)^2 \right) \drdtau 
\end{split}
\end{equation}
\begin{equation}
\begin{split}
\frac{\drm^2\varphi}{\drm\tau^2} = & \left( -\curlyh + 2 \Phi_{,\tau} + \Psi_{,\tau} \right) \dphidtau + \left( -\frac{2}{r} + 2 \Phi_{,r} + 2\Psi_{,r}  \right) \drdtau \dphidtau \\ 
&+ \left(\curlyh (1-2\Phi -2 \Psi ) -\Phi_{,\tau}\right) \left( \left(\drdtau\right)^2 + r^2 \left(\dphidtau\right)^2 \right) \dphidtau 
\end{split}
\end{equation}
\end{subequations}
\end{widetext}

In addition we have the ``mass-shell condition'' for particles with mass:
\begin{equation}
g_{\mu \nu} \frac{\drm x^{\mu}}{\drm s}  \frac{\drm x^{\nu}}{\drm s} = -1
\end{equation}
which is expressed, using our metric, as:\footnote{Again, we have set $\vartheta=\pi/2$, so $\drm \vartheta / \drm s =0$.}
\begin{multline}\label{eq:mass-shell}
-a^2 \left(\frac{\drm \tau}{\drm s}\right)^2 (1 + 2\Psi) 
+ a^2 \left(\frac{\drm r}{\drm s} \right)^2 (1 - 2\Phi) \\
+ a^2r^2\left(\frac{\drm \varphi}{\drm s}\right)^2 (1 - 2\Phi)
= -1
\end{multline}
The momentum of the particles now has a radial component,
\begin{equation}
p_r  = \frac{m (1-\Phi)\drdtau}{\sqrt{1 + 2 \Psi - (1-2\Phi) \left(\drdtau\right)^2 - (1-2\Phi) r^2 \left(\dphidtau\right)^2 }}
\end{equation}
and an angular one,
\begin{equation}
p_{\varphi} = \frac{m (1-\Phi) r \dphidtau}{\sqrt{1 + 2 \Psi - (1-2\Phi) \left(\drdtau\right)^2 - (1-2\Phi) r^2 \left(\dphidtau\right)^2 }}
\end{equation} 

Using the mass-shell condition (\ref{eq:mass-shell}), the definition of the conserved quantity $L$ (\ref{eq:dphids}), and (\ref{eq:d2ts}), we can extract the angular velocity of infinitesimal particles,
\begin{equation}
\dphidtau=\frac{L}{a r^2} \sqrt{ \frac{1+ 4\Phi + 2\Psi - (1+2\Phi)  \left(\drdtau\right)^2}{1 + \frac{L^2 (1+ 2 \Phi)}{a^2 r^2}}} \, ,
\end{equation} 
and eliminate it from the momenta equations: 
\begin{equation}
p_{r} = m \drdtau \frac{(1-\Phi) }{a r} \sqrt{\frac{a^2 r^2 + L^2 (1+2\Phi)}{1+ 2\Psi - (1- 2 \Phi)\left(\drdtau\right)^2 }} 
\end{equation} 
and 
\begin{equation}
p_{\varphi} = \frac{m L (1+\Phi)} {a r}. 
\end{equation}
Note that it is $L$ which is the conserved quantity, not $p_\varphi$, within our framework. With this setup, we can finally express the equation for evolution of shell particles (\ref{eq:d2rs}) as a system of two first-order equations:
\begin{multline}
\frac{\drm p_r}{\drm \tau} = 
(  \Phi_{,\tau} - \curlyh) p_r 
- \sqrt{m^2 + p_r^2 + p_{\varphi}^2} \Psi_{,r} \\
+ \left( \frac{1}{r} - \Phi_{,r} \right)\frac{p_{\varphi}^2(1+ \Phi + \Psi)}{\sqrt{m^2 + p_r^2 +p_{\varphi}^2}} 
\end{multline}
and 
\begin{equation}
\frac{\drm r}{\drm \tau} = 
\frac{p_r}{\sqrt{m^2 + p_r^2 + p_{\varphi}^2} } (1 + \Phi + \Psi ).
\end{equation}
These two equations describe the movement of shell particles and need to be solved numerically. A sanity check verifies that if we set $p_{\varphi}$ to zero, we recover the momentum evolution equation in the case without angular momentum, (\ref{eq:dpdtau-noang}).

With our new definitions of $p_r$ and $p_{\varphi}$, the energy-momentum tensor expresses as: 
\begin{equation}
T_0^0 = -\frac{1+3\Phi}{4 \pi r^2 a^3} \sum_n \delta(r - r_{(n)}) \left(\sqrt{m^2+p_r^2 + p_{\varphi}^2}\right)_{\!(n)}
\end{equation}
 and 
\begin{equation}
	\Pi_{rr}  = \frac{1+3\Phi}{4 \pi r^2 a^3} \sum_n \delta(r - r_{(n)}) \left(\frac{\frac{2}{3} p_r^2 - \frac{1}{3} p_{\varphi}^2}{ \sqrt{m^2+p_r^2 +p_{\varphi}^2} }\right)_{\!\!\!(n)}
\end{equation} 

The angular motion also gives rise to a transverse Doppler effect. This can be seen, e.g., from eq.~(\ref{eq:product_k_u}) being modified as
\begin{equation}
 g_{\mu\nu} k^\mu u^\nu = -k^0 a\left(1+\Psi\right)\frac{1}{m}\left[\sqrt{m^2+p_r^2+p_{\varphi}^2} - p_r\right] \, .
\end{equation}

In figure \ref{fig:orbits} we have plotted the radial trajectories of shells within balls that have uniform density, but non-zero and non-uniform angular momentum. Each panel corresponds to the same initial density state, but different initial states of angular momentum. As can be seen, our code is able to describe balls that collapse to a point; balls that first begin to collapse under gravity but then stabilise; balls where the effective pressure due to angular momentum perfectly balances gravitational attraction; balls that first expand due to effective pressure, but then stabilise; and finally, balls which are blown apart by pressure. Only in the first and last situations respectively will our code certainly break down, and even then it will survive until the weak field limit breaks down, or a particle leaves the box. In these plots, the balls begin only $50$ times larger than their Schwarzschild radii, therefore relativistic effects will not be negligible.

\section{Conclusion}

We have presented an N-body framework for spherically symmetric solutions valid in the weak-field regime of general relativity. We have primarily applied this framework in a cosmological context, expanding around an FLRW metric; however nothing forbids the application of the framework to other contexts.
Spherical symmetry was imposed in order to obtain an economic setup for numerical studies. We compared our code against two types of known exact solutions, Schwarzschild and Lema\^itre-Tolman-Bondi, and found good agreement. However, our scheme is suitable also for setups
where no exact solution is known, for instance when the fluid description of matter is not valid. 

Furthermore, we demonstrated
that the relativistic potentials are computed more accurately than in a Newtonian scheme. This feature will be useful for the
study of models which have exotic sources of stress-energy perturbations, such as dynamical dark energy or modified gravity.
On a related note, we stress that our scheme does not make any assumptions about the nature of perturbations apart from the
requirement that they give rise to weak gravitational fields only. This assumption breaks down, e.g., if a black hole is formed.

In order to avoid the collapse of an overdensity into a black hole, we have introduced a method to create a stable bound structure
supported by angular momentum. Such configurations may, in some sense, be more realistic proxies for cosmic
structures such as galaxy clusters, and can therefore be useful laboratories for studying gravity at these scales. They may also be useful for the study of weakly relativistic, compact bound objects that can form in the early universe, such as ultra compact mini-haloes, or for the early stages of the formation of primordial black holes.

\acknowledgments

We thank Peter Coles, Ruth Durrer, and Martin Kunz for useful suggestions, insights and comments.
This work was supported by the Swiss Society for Astrophysics and Astronomy through a Funding and Travel Award, sponsored by the Foundation MERAC
(Mobilising European Research in Astrophysics and Cosmology). SH acknowledges
support from the Science and Technology Facilities Council [grant number ST/L000652/1].
JA acknowledges support from the Swiss National Science Foundation.

\appendix

\section{Linear Relation between LTB Solution and Longitudinal Gauge}
\label{app:gauge}

In this appendix we give the linear gauge transformations which we use to set up initial conditions
corresponding to a given LTB solution. These relations are valid to the extent that the matching is done
at a time when linear perturbation theory can be applied. LTB solutions which do not allow a perturbative description
at any time are more difficult to translate into our gauge, and there may be cases where it is impossible.
We will not consider solutions of this type in this paper.

Our starting point is the LTB line element given in eq.~(\ref{eq:LTBmetric}). At the initial time $t_{\mathrm{in}}$
where we will do the matching, we will rescale the coordinate $r$ such that, at that time, $R(t_{\mathrm{in}}, r) = a(t_{\mathrm{in}}) r$.
The (time independent) gravitational mass function can then simply be obtained as
\begin{eqnarray}
 2 M(r) &=& 8 \pi G a^3(t_{\mathrm{in}}) \int_0^r \tilde{r}^2 \rho(t_{\mathrm{in}}, \tilde{r}) \drm\tilde{r}\nonumber\\
 &=& 3 H^2 a^3 \int_0^r \tilde{r}^2 \left(1+\delta(t_{\mathrm{in}}, \tilde{r})\right) \drm\tilde{r} \, .
\end{eqnarray}
In this work we are interested in setups where the metric is FLRW everywhere except for a finite spherical region.
This can be achieved by choosing a ``compensated'' density profile such that the mass function becomes the
one of FLRW, $M(r) = H^2 a^3 r^3 / 2$, at the boundary of the region\footnote{Note that $H^2 a^3$ is independent of time in a matter dominated universe.}. We will choose a particularly simple profile, given by
\begin{equation}
 \delta(t_{\mathrm{in}}, r) = \begin{cases}
                           \delta_1  & r<r_1 \\
                           \delta_2  & r_1<r<r_2 \\
                           0 & r > r_2
                          \end{cases} \, ,
\end{equation}
where $\delta_2 = -\delta_1 r_1^3 / \left(r_2^3 - r_1^3\right)$ gives the correct matching to FLRW as can be seen
by inspecting the resulting mass function.

Next we will use the parametric expression for the exact LTB solution in order to determine the metric function $E(r)$.
Let us consider $E(r) < 0$, the opposite case is analogous. The parametric expressions are
\begin{subequations}
\begin{eqnarray}
 R(t,r) &=& -\frac{M(r)}{2 E(r)}\left(1-\cos\eta\right) \, ,\\
 \left(\eta - \sin\eta\right)^{2/3} &=& -\frac{2 E(r)}{M^{2/3}(r)} t^{2/3} \, .
\end{eqnarray}
\end{subequations}
While the parameter $\eta$ cannot be eliminated in closed form, it is possible to do so perturbatively for small $\eta$,
i.e.\ at early time. We find that
\begin{equation}
 E(r) = -\frac{5}{6} r^2 H^2(t_{\mathrm{in}}) a^2(t_{\mathrm{in}}) f(r) \, ,
\end{equation}
where
\begin{equation}
 f(r) = \begin{cases}
         \delta_1 & \mathrm{if}~ r<r_1\\
         \delta_2 + \frac{r_1^3}{r^3}\left(\delta_1-\delta_2\right) & \mathrm{if}~ r_1<r<r_2\\
         0 & \mathrm{if}~ r>r_2
        \end{cases} \, .
\end{equation}
In the linear regime we also have
\begin{equation}
 R(t,r) = a(t) r \left[1 + \frac{1}{3} f(r) \left(1 - \frac{a(t)}{a(t_{\mathrm{in}})}\right)\right] \, .
\end{equation}
Let us now write the line element in a convenient perturbative notation,
\begin{subequations}
\begin{eqnarray}
 R^2(t, r) &=& a^2(t) r^2 \left[1 + 2 b(t, r)\right] \, ,\\
 \frac{\left[R_{,r}(t,r)\right]^2}{1+2E(r)} &=& a^2(t) \left[1+2y(t,r)\right] \, ,
\end{eqnarray}
\end{subequations}
which implies
\begin{subequations}
 \begin{eqnarray}
  b(t,r) &=& \frac{1}{3} f(r) \left(1 - \frac{a(t)}{a(t_{\mathrm{in}})}\right) \, ,\\
  y(t,r) &=& b(t,r) + r b_{,r}(t,r) - E(r) \, .
 \end{eqnarray}
\end{subequations}
We can now work out the linear gauge transformation from synchronous to longitudinal gauge (see also \cite{Adamek:2014qja}). A straightforward calculation shows
\begin{widetext}
\begin{subequations}
 \begin{eqnarray}
  r \frac{\partial}{\partial r} \left[\frac{1}{r} \Phi_{,r}(t,r)\right] &=& -b_{,rr}(t,r) + H(t) a^2(t) \left[y_{,t}(t,r) - b_{,t}(t,r)\right] - \frac{2}{r^2}\left[y(t,r) - b(t,r)\right] + \frac{1}{r} y_{,r}(t,r) \, , \\
  r \frac{\partial}{\partial r} \left[\frac{1}{r} \Psi_{,r}(t,r)\right] &=& 2 H(t) a^2(t) \left[b_{,t}(t,r) - y_{,t}(t,r)\right] + a^2(t) \left[b_{,tt}(t,r) - y_{,tt}(t,r)\right] \, .
 \end{eqnarray}
\end{subequations}
\end{widetext}
Since the combination $H^2 a^3$ is independent of time in a matter dominated universe we can evaluate it at $t=t_{\mathrm{in}}$ and find
\begin{multline}
 r \frac{\partial}{\partial r} \left[\frac{1}{r} \Phi_{,r}(t,r)\right] = r \frac{\partial}{\partial r} \left[\frac{1}{r} \Psi_{,r}(t,r)\right] \\ = \frac{1}{2} H^2(t_{\mathrm{in}}) a^2(t_{\mathrm{in}}) r f'(r) \, ,
\end{multline}
which can be integrated twice to obtain $\Phi$ and $\Psi$. The constants of integration should be chosen such that we can match
smoothly to FLRW at $r=r_2$. In other words, we require $\Phi\vert_{r=r_2} = \Psi\vert_{r=r_2} = \Phi_{,r}\vert_{r=r_2} = \Psi_{,r}\vert_{r=r_2} = 0$. The corresponding solutions are
\begin{equation}
 \Phi(t,r) = \Psi(t,r) = \frac{1}{2} H^2(t_{\mathrm{in}}) a^2(t_{\mathrm{in}}) \int_{r_2}^r \tilde{r} f(\tilde{r}) \drm\tilde{r} \, .
\end{equation}

\section{Initial Particle Data}
\label{app:ic}

Given a linear solution for $\Phi$, $\Psi$ which specifies the initial conditions, we can work out the initial particle configuration.
To this end, we linearize eq.~(\ref{eq:Einstein1}),
\begin{equation}
\label{eq:linEinstein1}
 \Phi_{,rr} + \frac{2}{r} \Phi_{,r} - 3 \curlyh^2 \Phi = - 4 \pi G a^2 \delta T_0^0 \, ,
\end{equation}
where we have used that $\Phi_{,t} = \chi = 0$ at linear order.

The aim of this section is to construct a linear displacement field $\delta r (r)$ which specifies the initial particle positions
$r_{(n)}(t_{\mathrm{in}}) = r_{(n)}^0 + \delta r(r_{(n)}^0)$ as infinitesimal displacements from a homogeneous distribution $r_{(n)}^0$.
Expanding eq.~(\ref{eq:T00}) to linear order we find
\begin{equation}
\label{eq:linT00}
 \delta T_0^0 = \bar{T}_0^0 \left(3 \Phi - \delta r_{,r} - \frac{2}{r} \delta r\right) \, .
\end{equation}
To see this, take the continuum limit of the particle sum,
\begin{multline}
 T_0^0 = -\frac{1+3\Phi}{4 \pi r^2 a^3} \sum_n m_{(n)} \delta(r-r_{(n)})\\ \rightarrow -\frac{1+3\Phi}{4 \pi r^2 a^3} \int \bar{f}(r_{(n)}^0) \delta(r-r_{(n)}) \drm r_{(n)}^0 \, ,
\end{multline}
with a distribution function $\bar{f}(r_{(n)}^0) \propto (r_{(n)}^0)^2$ corresponding to the homogeneous distribution. Next, change
the integration variable to $r_{(n)}$ to obtain eq.~(\ref{eq:linT00}). Inserting into eq.~(\ref{eq:linEinstein1}) and using
eq.~(\ref{eq:Friedmann}), the solution for the displacement is found to be
\begin{equation}
 \delta r = \frac{5}{r^2} \int_0^r \tilde{r}^2 \Phi(t_{\mathrm{in}}, \tilde{r}) \drm\tilde{r} - \frac{2}{3 \curlyh^2} \Phi_{,r} \, .
\end{equation}
Here the constant of integration is fixed by requiring regularity at the origin, which implies $\delta r\vert_{r=0} = 0$.

The initial particle velocities can be obtained simply by taking the time derivative of above equation,
\begin{equation}
 \left.\frac{\drm r}{\drm \tau}\right|_{t_{\mathrm{in}}} = \delta r_{,\tau} = -\frac{2}{3 \curlyh} \Phi_{,r} \, ,
\end{equation}
where we used $\curlyh' = -\curlyh^2 / 2$ in a matter dominated universe.

\section{Particle-Mesh Scheme for Spherical Coordinates}
\label{app:pm}

Standard particle-mesh schemes \cite{HockneyEastwood} usually employ a Cartesian mesh which means that a few modifications are required
in order to make them fit for our purpose. Our mesh will have a uniform resolution in $r$, the radial coordinate, meaning that the volume of
the cells increases as one moves outwards from the center. The mass resolution can be set independently by changing the number of particles
per cell -- a number which can also depend on radius and should be chosen according to the problem at hand.

The stress-energy tensor on the grid is computed by means of a particle-to-mesh projection. It is constructed by smearing out each particle
over a finite radial interval and then determining the fraction of its mass within each cell. Explicitly, we replace
\begin{equation}
 \frac{\delta(r - r_{(n)})}{4 \pi r^2} \underset{r=r_i}{\rightarrow} \frac{w(r_i - r_{(n)})}{V_i} \, ,
\end{equation}
where $r_i$ denotes the center of the $i$th cell, $V_i$ is the cell's volume, $r_{(n)}$ is the position of the $n$th particle, and $w$ is a
weight function which depends on the smearing. We use triangular-shaped particles where
\begin{equation}
 w(r_i - r_{(n)}) = \int {\sqcap}(r - r_i) {\wedge}(r - r_{(n)}) \drm r \, ,
\end{equation}
with
\begin{equation}
 {\sqcap}(\Delta r) = \begin{cases}
                  1 & \mathrm{if}~ -\frac{\drm r}{2} < \Delta r < \frac{\drm r}{2} \\
                  0 & \mathrm{otherwise}
                 \end{cases} \, ,
\end{equation}
and
\begin{equation}
 {\wedge}(\Delta r) = \begin{cases}
                      1+\frac{\Delta r}{\drm r} & \mathrm{if}~ -\drm r < \Delta r < 0 \\
                      1-\frac{\Delta r}{\drm r} & \mathrm{if}~ 0 < \Delta r < \drm r \\
                      0 & \mathrm{otherwise}
                     \end{cases} \, .
\end{equation}
Here and in the following, $\drm r$ denotes the grid unit. Pictorially, ${\sqcap}$ characterizes the footprint of the cell (an interval of width $\drm r$ centered at $r=r_i$), whereas
${\wedge}$ characterizes the shape of the particles (its mass distribution along the radial coordinate).

Our grid cells are centered at $r_i = i \drm r$, with $i = 0, 1, \ldots$. The volume of the $i$th cell is computed as
\begin{multline}
 V_i = \frac{4 \pi}{3} \left(r_i + \frac{\drm r}{2}\right)^3 - \frac{4 \pi}{3} \left(r_i - \frac{\drm r}{2}\right)^3 \\= 4 \pi \left(i^2 + \frac{1}{12}\right) \drm r^3
\end{multline}
for $i>0$, and
\begin{equation}
 V_0 = \frac{4 \pi}{3} \left(\frac{\drm r}{2}\right)^3 = \frac{4 \pi}{24} \drm r^3
\end{equation}
for the cell at the origin. Contributions which would be projected at negative radius are simply folded back onto the positive axis.

Next, we want to construct a particle distribution which would correspond to a homogeneous universe. A perturbed distribution can then be obtained by acting with an
infinitesimal displacement as explained in appendix \ref{app:ic}. Our homogeneous distribution will be constructed in a way as to minimize discretization issues.
For simplicity, let us discuss a setup where we have one particle per grid cell (mass resolution can be increased by subdividing particles). If we choose initial particle
positions as $r_{(n)} = (n + \frac{1}{2}) \drm r$, with $n = 0, 1, \ldots$, one can recursively construct the mass assignment for each particle which would lead to
an exactly uniform projected density under the above projection method:
\begin{equation}
 m_{(n)} = 4 \pi \left(\frac{1}{12} + n + n^2\right) \drm r^3 a^3 \bar{\rho}
\end{equation}
Here, $\bar{\rho} = -\bar{T}^0_0$ is the homogeneous density of the background FLRW model. Evidently, at very large radius $r_{(n)} \gg \drm r$ this expression asymptotes to the correct continuum limit $m_{(n)} = 4 \pi r_{(n)}^2 \drm r a^3 \bar{\rho}$. The corrections which come in at small radii are chosen to compensate for discretization effects, such that the projected density remains exactly homogeneous,
\begin{equation}
 \frac{1}{a^3} \sum_n \frac{w(r_i - r_{(n)})}{V_i} m_{(n)} = \bar{\rho} \qquad \forall i \, .
\end{equation}

The weight function $w$ can also be used in order to interpolate grid-based quantities (fields, gradients of fields etc.) to the positions of the particles. This
is necessary for the integration of the geodesic equations. For instance, in order to interpolate $\Phi$, we would define
\begin{equation}
 \Phi(r_{(n)}) = \sum_i \Phi(r_i) w(r_i - r_{(n)}) \, .
\end{equation}
Note that the sum is now taken over the grid points. Similarly, we can interpolate a gradient as
\begin{equation}
 \Psi_{,r}(r_{(n)}) = \sum_i \frac{\Psi(r_{i+1}) - \Psi(r_i)}{\drm r} w(r_i + \frac{1}{2}\drm r - r_{(n)}) \, ,
\end{equation}
based on a standard one-sided two-point gradient which naturally sits at half-integer grid units.

\bibliographystyle{utcaps}
\bibliography{spherical}

\end{document}